\documentstyle[aps]{revtex}

\begin{document}
\draft
\baselineskip=12pt

\title{The quantum mechanics of particle-correlation
           measurements in high-energy heavy-ion collisions}
\author{Alejandro Ayala,$^{\rm a}$ Gordon Baym,$^{\rm b}$, and
James L. Popp$^{\rm b,c}$}

\address{$^{\rm a}$ Instituto de Ciencias Nucleares,
Universidad Nacional Aut\'onoma de M\'exico,
Aptdo.\ Postal 70-543, M\'exico D.F. 04510\\
$^{\rm b}$ Department of Physics, University of Illinois at
Urbana-Champaign, 1110 W. Green St., Urbana, IL 61801 \\
$^{\rm c}$ Department of Physics, New York University,
4 Washington Place, New York, NY 10003}
\date{\today}
\maketitle
\begin{abstract}

    The Hanbury Brown--Twiss (HBT) effect in two-particle correlations is a
fundamental wave phenomenon that occurs at the sensitive elements of
detectors; it is one of the few processes in elementary particle detection
that depends on the wave mechanics of the produced particles.  We analyze
here, within a quantum mechanical framework for computing correlations among
high-energy particles, how particle detectors produce the HBT effect.  We
focus on the role played by the wave functions of particles created in
collisions and the sensitivity of the HBT effect to the arrival times of pairs
at the detectors, and show that the two detector elements give an enhanced
signal when the single-particle wave functions of the detected particles
overlap at both elements within the characteristic atomic transition time of
the elements.  The measured pair correlation function is reduced when the
delay in arrival times between pairs at the detectors is of order of or larger
than the transition time.

\end{abstract}


\section{INTRODUCTION}

    The Hanbury Brown--Twiss (HBT) effect in nuclear and particle
physics~\cite{PION:HBT1,PION:HBT2,PION:HBT3,PION:HBT4,PION:HBT5,PION:CURRENTS}
is the enhancement at small relative momentum in the
probability for observing pairs of identical bosons drawn from the same
collision event, compared to that for pairs chosen from different events.
Underlying the correlations between the fluctuations at two nearby but
separate detectors is the symmetry of the wave function of a system of
identical bosons under interchange of any two particles.  HBT particle
interferometry is in a special class of high energy physics experiments that,
like kaon regeneration and neutrino oscillation searches, study an effect
basic to quantum mechanics:  amplitude interference.  However, while, e.g.,
kaon regeneration experiments examine how weak-interactions affect the {\it
internal} degrees of freedom of kaons, HBT interferometry probes the many-body
{\it spatial} wave function of the final state particles.  The effect, as we
emphasize here, is fundamentally a {\it wave phenomenon} manifested {\it at
the detectors}.  Indeed, as is well understood in terms of classical wave
mechanics, HBT intensity-intensity correlations may be observed with any type
of wave satisfying the superposition principle (e.g., sound waves) with at
least two independent incoherent sources and two detectors.

    In this paper we analyze from a quantum mechanical point of view the
physics of correlations between identical particles, and how the detectors in
high-energy physics experiments probe the many-particle wave functions in
space and time.  While the spatial requirements on the detector separations
for measuring an HBT effect are well understood, the temporal requirements are
less well studied.  We may ask, what is the maximum separation in time of
arrival of two identical particles at the detectors that will still yield an
HBT effect between the two particles?  It is commonly assumed in collision
experiments that interference happens only between particles produced in the
same event, not between particles from different events.  One could imagine
that this occurs because the correlations are built in at the time of the
event; however since one detects interferometry between photons from the
opposite edges of a star, where there can clearly be no correlation between
the emission processes, this is certainly not the case.  Underlying the
assumption of lack of interferometry between particles from different events
is the existence of a timescale for detection of interferometry, which is
exceeded by successive events.  As we show here, the crucial timescale for HBT
correlations is that inherent in the detection process; the maximum value of
the measured HBT pair correlation function is reduced when pairs of waves
arrive at the detector atoms with delay times comparable to or larger than the
characteristic transition times of the atoms in the detector.

    Considering the wave mechanics of the emitted particles in space and time
also enables us to understand more fully questions such as:  How does the
exchange symmetry of the many-particle wave function lead to detection of
momentum correlations?  How does the wave packet nature of the single-particle
states affect measurements of the HBT effect?  For example, how does the
spreading of wave packets affect the measurement of correlations?

    Part of the motivation to examine the role of the interplay of the time
scales of the pion field and the detection process in particle interferometry
experiments is the clear illustration in the original studies of photon
correlations of how detection times affect the size of the maximum value of
the measured two-photon correlation function
\cite{R:HanburyBrown,R:HanburyBrown24}.  As Purcell~\cite{HBT:Purcell} pointed
out, for chaotic stationary sources the maximum height of this function above
unity is approximately the ratio of the coherence time of the light to the
observational binning time.  Hanbury Brown and Twiss's measurements on the
star Sirius~\cite{HBT:Sirius} produced a maximum correlation signal $\sim
10^{-6}$, in contrast to data on like-pion pair correlations gathered to date,
with final-state interaction corrections included, which indicate that the
two-pion correlation function, $C_2(|{\vec q}\,|)$, rises approximately (but
not unambiguously) to two in the lowest bin in relative momentum.

    Four time scales are basic in a measurement of the HBT effect.  The first
is the {\it lifetime} of the source, $\tau_{\rm s}$, over which it emits
particles; in a high-energy collision the production region exists for $\sim$
1-10 fm/c [$\sim 0.3-3 \times 10^{-23}$ sec] in its own rest frame.  Second is
the {\it coherence time} of the source, $\tau_{\rm c}$, the characteristic
time for each elementary radiator to emit a wave.  Generally, the formation
time for a pion or other particle depends on its energy; the characteristic
lifetime of elementary pion sources $\tau_{\rm c}$ is $\raisebox{-0.1
cm}{$\stackrel{<}{\sim}$}$ 1 fm/c, and $\tau_{\rm c}
\raisebox{-0.75ex}{$\stackrel{<}{\sim}$}\tau_{\rm s}$.  The third scale is the
{\it atomic transition time}, $\tau_{\rm atomic}$, of the energy-absorbing
material of the detector, the time over which a mobile electric charge is
created in the detector; on this time scale the detector atoms ``do quantum
mechanics'' on the incoming particle waves, i.e., the atoms are sensitive to
the amplitudes and phases of the waves.  For ionization of a gas atom,
$\tau_{\rm atomic} \sim \hbar/10\,{\rm eV} \sim 10^{-16}\,$sec.  The final
scale is the {\it exposure time}, $\tau_{\rm exp}$, the interval between
observations of the state of the detector atoms.  The time of an accurate
momentum measurement on a relativistic charged particle in a magnetic
spectrometer of length $\ell_{\rm spec}$ is $\sim \ell_{\rm spec}/c$, which
for a typical length scale of $10\,$m gives a characteristic time to
measure the momentum $\tau_{\rm exp} \sim 3\times 10^{-8}\,$sec.  A more
familiar but less relevant measurement time scale is the detector {\it
resolution time}, $\tau_{\rm res}$, literally, the minimum time to detect an
electronic signal, which includes the time it takes to collect and amplify the
initial electric charge.  A typical resolution time for a wire chamber,
essentially the rise time of the voltage pulse produced by an electron
avalanche on an anode wire, is on a nanosecond scale, $\tau_{\rm res} \sim
10^{-9}\,$sec.  As we show in Sect.~III, $\tau_{\rm atomic}$, rather than
$\tau_{\rm res}$, is the important time scale over which HBT correlations are
detected.

    This paper is organized as follows:  In Section~II, we develop a framework
for understanding the temporal structure in correlations between quantum
mechanical particles created in high-energy experiments.  To be specific, we
describe pions, but our results hold for other particles as well.  In
Section~III, we analyze the features of relativistic wave packets important to
HBT, and compute the detection probability for pairs directly-produced in a
heavy-ion collision and then for those from direct production plus resonance
decay.  In Section~IV, we summarize our results and conclusions.  In the
Appendices, we describe the details of the quantum mechanics of particle
detection.

\section{DESCRIPTION OF THE PROBLEM}

    One and two pion measurements on a multiparticle system are described by
the single-pion and two-pion density matrices for particles of given charge:
$\langle \phi^{\dagger}(x_1 ) \phi(x_2 )\rangle$ and $\langle
\phi^{\dagger}(x_1 ) \phi^{\dagger}(x_2 ) \phi(x_3 ) \phi(x_4 ) \rangle$,
where $\phi(x)$ is the part of the (Heisenberg representation) pion field
operator that destroys particles of the given charge, and the brackets
indicate an ensemble average over the states of the colliding nuclei.  The
pion-pair correlation function, $C_{2}({\vec q}\,)$, which depends directly on
these two functions, is measured as the ratio of the pion pair distribution to
the separate single distributions:
\begin{equation}
C_2({\vec q}\,) = \frac{\{ d^6 n_2 /
dp^3 d{p'}^{\raisebox{-.35ex}{\mbox{\scriptsize 3}}}\}}
{\{ d^3n_1 / dp^3 \;
d^3n_1 / d{p'}^{\raisebox{-.35ex}{\mbox{\scriptsize 3}}}\}},
\label{eq:corfuncdef}
\end{equation}
where ${\vec q} = ({\vec p} - {\vec p}\,')/2$.  The braces in the
numerator denote an average over an ensemble of pairs drawn from the same
event and in the denominator they denote an average over an ensemble of pairs
drawn from different events.  The single-pion momentum distribution $d^3n_1 /
dp^3$ is given in terms of the plane-wave momentum state creation and
annihilation operators $a^{\dagger}_{\vec p}$ and $a^{\phantom{\dagger}}_{\vec
p}$ by
\begin{equation}
\frac{d^3n_1}{dp^3} =
\langle a^{\dagger}_{\vec p}\, a^{\phantom{\dagger}}_{\vec p} \rangle =
2\varepsilon_{p} \int\!
d^3r_{1}\, d^3r_{2}\; e^{-ip\cdot (x_1 - x_2 )} \,
\langle \phi^{\dagger}(x_1 ) \phi(x_2 ) \rangle \, ,
\label{eq:spectra}
\end{equation}
with $\varepsilon_p = ({\vec p}^{\, 2} + m^2)^{1/2}$ and $p\cdot{x} \equiv
\omega t - {\vec p}\cdot{\vec r}$, where $\omega$ is the particle energy
and $\vec p$ is the momentum.  Note that the explicit dependence on $t_1$
and $t_2$ in the phase factor of Eq.~(\ref{eq:spectra}) is canceled by
the time-dependence of the single-pion density matrix, as later
shown in Eq.~(\ref{eq:spec}).  The momentum distribution of pairs is
given similarly by
\begin{equation}
\frac{d^6 n_2}{dp^3 d{p'}^{\raisebox{-.35ex}{\mbox{\scriptsize 3}}}}
= \langle a^{\dagger}_{\vec p} \, a^{\dagger}_{{\vec p}\, '} \,
a^{\phantom{\dagger}}_{{\vec p}\, '} \, a^{\phantom{\dagger}}_{\vec p} \rangle
= 4 \varepsilon_{p} \varepsilon_{p'} \int\! d^3r_1 \cdots d^3r_4\;
e^{-ip\cdot (x_1 - x_4 )} e^{-ip'\cdot (x_2 - x_3 )} \;
\langle \phi^{\dagger}(x_1 ) \phi^{\dagger}(x_2 ) \phi(x_3 ) \phi(x_4 )
\rangle \, .
\label{eq:spectra2}
\end{equation}

    A detailed description of the pion wave functions that emerge from the
production region of a heavy-ion collision requires knowledge of the evolution
and geometry of the source and how field excitations of quarks, gluons,
nucleons, etc. form the currents that radiate pions.  However, these details
are unimportant for our present purpose of understanding how we detect HBT.
We retain only the essential features of the pion production process, in
particular its characteristic length and time scales.  In high-energy
collisions, pions are produced in bremsstrahlung-like processes, in states
similar to momentum wave packet states.  Final-state interactions,
particularly Coulomb and strong interactions, severely influence particle
state evolution; also, collisions between pions and air molecules, detector
materials, and other target nuclei can significantly alter $C_{2}$; in
particular, multiple scatterings of particles after they leave the source
redistribute pion tracks thereby reducing the size of the measured correlation
in each bin in relative momentum and decreasing the effective size of the
source measured in HBT experiments~\cite{BAYM:zak,ABJPV:air}.  Thus in order
to focus on the wave mechanics of detecting HBT, we assume here that after
their last strong interaction the wave packets propagate in vacuum.

    The production of charged pions and their propagation to the detectors is
described by the Klein-Gordon equation for the charged pseudoscalar field,
which relates the pion field, $\phi$, to the source, $J(x)$, of the pion
field at the {\it last} strong interaction:
\begin{equation}
\left(
\partial^2 / \partial t^2 - \nabla^2  + m_{\pi}^2 \right) \phi(x) = - J(x),
\label{eq:KGsource}
\end{equation}
where $J^\dagger$ is the part of the current operator that emits pions of
the given charge.  Thus
\begin{equation}
\phi(x) =  \int\! dx' \; D_{\rm ret}(x-x') J(x') \, ,
\label{eq:wave1}
\end{equation}
where the free-field retarded Green's function, $D_{\rm ret}(x-x')$, satisfies
\begin{equation}
\left(\partial^2 / \partial t^2 - \nabla^2  + m^2_{\pi} \right)
D_{\rm ret}(x-x') = - \delta^{(4)}(x-x' ),
\end{equation}
and vanishes for $t < t'$; the integrations are over all space and time.
The Green's function has the representation
\begin{equation}
D_{\rm ret}(x-x') = \int\! \frac{d^4 k}{(2\pi )^4} \;
\frac{e^{-ik\cdot(x-x')}}
{(\omega +i\epsilon)^2 - \varepsilon^2_{k}} \, ,
\label{eq:niretGF}
\end{equation}
where $\epsilon$ is a positive infinitesimal and $\omega=k^0$.
Integrating over $\omega$
for $t>t'$, we find that the created pion field is given by
\begin{equation}
\phi(x) = \int\!\frac{d^{3}k}{2i\varepsilon_{k}(2\pi)^3} \;
\int^{t}_{-\infty}\! dt' \int\! d^{3}r' \; J(x') \, e^{-ik\cdot (x-x')},
\label{eq:intpionfld}
\end{equation}
with $k$ on-shell.  Since one measures at $t$ much larger than the source
lifetime we extend the upper limit in the time integral in
Eq.~(\ref{eq:intpionfld}) to $+\infty$ and write
\begin{equation}
\phi(x) = \int\! \frac{d^3 k}{2i\varepsilon_{k}(2\pi )^3 }\;
e^{-ik\cdot x} \, J(k) \, ,
\label{eq:piplusff}
\end{equation}
where $J(k)$ is the Fourier transform in space and time of the pion source
current operator.  Equations~(\ref{eq:spectra}) and (\ref{eq:spectra2}) imply
that the Lorentz-invariant one pion momentum distribution is given by
\begin{equation}
2\varepsilon_p \frac{d^3 n_1 } {dp^{\raisebox{.4ex}{\mbox{\scriptsize 3}}}}
= \int\! dx_1 dx_2\; e^{-ip\cdot(x_1 - x_2)}
\langle J^{\dagger}(x_1) J(x_2) \rangle
\label{eq:spec}
\end{equation}
and the two pion momentum distribution by
\begin{equation}
4\varepsilon_p \varepsilon_{p'} \frac{d^6 n_2}
{dp^{\raisebox{.4ex}{\mbox{\scriptsize 3}}} d{p'}^3}
= \int\! dx_1 dx_2 dx_3 dx_4\; e^{-ip\cdot(x_1 - x_4) -ip'\cdot(x_2 - x_3)}
\langle J^{\dagger}(x_1) J^{\dagger}(x_2) J(x_3) J(x_4) \rangle;
\label{eq:spec2}
\end{equation}
Eqs.~(\ref{eq:spec}) and (\ref{eq:spec2}) show how the HBT correlation
function directly probes the correlation functions of the source, a point
of view initially introduced in Ref.~\cite{PION:CURRENTS}.

    The HBT effect arises only when single particles are produced in {\it
mixed} quantum states, that is, in an ensemble or statistical mixture of
single-particle states.  (See Ref.~ \cite{MS:COLL} for further discussion of
the multiparticle states in HBT.)  When particles are produced {\it completely
independently}, e.g., by a thermal source, the source currents factorize as
\begin{equation}
\langle J^{\dagger}(x_1 ) J^{\dagger}(x_2 ) J(x_3 ) J(x_4 ) \rangle
= \langle J^{\dagger}(x_1 ) J(x_4 ) \rangle
\langle J^{\dagger}(x_2 ) J(x_3 ) \rangle
+ \langle J^{\dagger}(x_1 ) J(x_3 ) \rangle \langle
J^{\dagger}(x_2 ) J(x_4 ) \rangle.
\label{eq:fact}
\end{equation}
The pion correlation functions similarly factorize, and the momentum
distribution of pion pairs~\cite{Pratt} is
\begin{equation}
\frac{d^6 n_2} {dp^{\raisebox{.4ex}{\mbox{\scriptsize 3}}} d{p'}^3}
= \frac{d^3 n_1 } {dp^{\raisebox{.4ex}{\mbox{\scriptsize 3}}}}
\frac{d^3 n_1 }{d{p'}^3}
+ \frac{1} {4\varepsilon_{\vec p}\, \varepsilon_{{\vec p}\,'}}\; \big|
\langle J^{\dagger}(p) J(p') \rangle \big|^2;
\label{eq:pra}
\end{equation}
HBT interferometry seeks to measure the second term in
Eq.~(\ref{eq:fact}).  The key physical mechanism that leads to the
factorization of currents is the loss of phase correlations among the
elementary sources, which is expected to occur in heavy-ion collisions through
the considerable rescattering of pions in the production region.\footnote{In
the absence of rescattering the statistics of the constituents of the currents
can destroy this factorization; a simple example is the correlations among
pions radiated by a weakly-interacting gas of
nucleons~\cite{BAYM:zak,ABP:curcor}; see also~\cite{MS:COLL}.} The HBT effect
is maximum, for incoherent emission.

    By contrast, when particles are produced {\it completely coherently},
e.g., as in an atom laser beam extracted from a Bose-Einstein
condensate~\cite{W:ketterle} or by an ideal chiral condensate~\cite{Bj}, the
source currents factorize as
\begin{equation}
\langle J^{\dagger}(x_1 ) J^{\dagger}(x_2 ) J(x_3 ) J(x_4 ) \rangle
= \langle J^{\dagger}(x_1 ) \rangle
\langle J^{\dagger}(x_2 ) \rangle \langle J(x_3 ) \rangle
\langle J(x_4 ) \rangle \, .
\label{eq:coherent}
\end{equation}
In this case, single particles are produced in a {\it pure} quantum state;
the momentum distribution of pion pairs is the product of two single-pion
momentum distributions, and the HBT effect would be absent.

\section{MEASURING CORRELATION FUNCTIONS}

    We turn now to the physics of momentum measurements.  How ionization
chambers (and many other detectors used in high-energy physics) track
well-localized particles is understood from semi-classical ideas; however, the
response of such detectors to a particle whose wave function is not
well-localized in space and time necessarily requires a quantum mechanical
description (a problem of longstanding interest, e.g., from Ref.~\cite{Mott}
to \cite{MS:hydro}.)  HBT correlations are the direct result of the
interactions of the {\it many-pion} wave function with the electrons of the
energy absorbing material of the detector.  Computing the probability of
measuring the momentum of a fast charged particle is a problem in multiple
scattering, where we must compute the probability of measuring a particle
track through a spectrometer system.  The first interaction selects out the
direction of the momentum, and subsequent interactions with more atoms and a
magnetic field select out the magnitude of the momentum.  We define the {\it
single-pion momentum measurement probability} $P^{\rm a}_{\pi}({\vec k}\,)$ as
the probability for a pion to ionize a detector gas atom at some location
${\vec a}$, say the first atom along a track, and undergo a transition to a
plane-wave momentum state ${\vec k}$.  Similarly, we define the {\it pion-pair
momentum measurement probability} $P^{\rm ab}_{\pi\pi}({\vec k}\, , {\vec
k}'\, )$ as the probability for each pion to ionize one atom and emerge in a
plane wave state, one atom at ${\vec a}$ with a pion in the final state ${\vec
k}$ and one atom at ${\vec b}$ with a pion in the final state ${\vec k}'$.

    As derived in Appendix A, the crucial function describing the
response~\cite{RJG:LH1964} of the detector in measuring a particle of momentum
${\vec k}$ -- a ``momentum measurement'' as defined above -- is the {\it
spectrometer function},
\begin{equation}
{\cal S}^{\rm a}_{\vec k}(x_1 \, x_2 ) \equiv
e^4 \, \psi^{\phantom{*}}_{\vec k} (x_1 ) \,
\langle {\stackrel{\,\sim}{j}}^{\dagger}_{\!\rm a}(x_1 )
{\stackrel{\,\sim}{j}}^{\phantom{\dagger}}_{\!\rm a}(x_2 ) \rangle \,
\psi^{*}_{\vec k} (x_2 ),
\label{eq:specfunc}
\end{equation}
where ${\stackrel{\,\sim}{j}}_{\!\rm a}(x)$ is the effective
electromagnetic current operator for the atomic electrons, Eq.
(\ref{eq:pseudxo}), and $\psi_{\vec k}$ is the final pion plane-wave
momentum state wave function.
As shown by detailed calculation in Appendices~A and B, the momentum
measurement probabilities are given by the pion correlation function as
filtered by the ``spectrometer function," i.e., the overlap in space and
time of the single and two-pion correlation functions (for given final pion
states) with the correlation function of the effective electron currents
in the detector atoms,
\begin{equation}
P^{\rm a}_{\pi}({\vec k}\,) = \int\! dx_1 dx_2 \;
{\cal S}^{\rm a}_{\vec k}(x_1 \, x_2 )\;
\langle \phi^{\dagger}(x_1 ) \phi(x_2 ) \rangle,
\label{eq:piexact0}
\end{equation}
and
\begin{equation}
P^{\rm ab}_{\pi\pi} ({\vec k}\, , {\vec k}'\, )=\int\! dx_1 dx_2 dx_3 dx_4 \;
{\cal S}^{\rm a}_{\vec k}(x_2 \, x_3 ) \;
\langle \phi^{\dagger}(x_1 ) \phi^{\dagger}(x_2 )
\phi(x_3 ) \phi(x_4 ) \rangle \;
{\cal S}^{\rm b}_{{\vec k}'}(x_1 \, x_4 ) \, .
\label{eq:doubdet2piprob0}
\end{equation}

    When the source term factorizes as in Eq.~(\ref{eq:fact}), we obtain the
pair momentum detection probability as a sum of two terms,
\begin{equation}
P^{\rm ab}_{\pi\pi} ({\vec k}\, , {\vec k}'\, ) =
D^{\rm ab}_{\pi\pi} ({\vec k}\, , {\vec k}'\, ) +
E^{\rm ab}_{\pi\pi} ({\vec k}\, , {\vec k}'\, ) \, ,
\label{eq:doubdet2pifact}
\end{equation}
where the {\em direct} term is
\begin{eqnarray}
D^{\rm ab}_{\pi\pi} ({\vec k}\, , {\vec k}'\, )
& = &
\int\! dx_1 dx_2 dx_3 dx_4 \;
{\cal S}^{\rm a}_{\vec k}(x_2 \, x_3 )\,
{\cal S}^{\rm b}_{{\vec k}'}(x_1 \, x_4 )\;
\langle \phi^{\dagger}(x_2 ) {\phi}(x_3 ) \rangle
\langle \phi^{\dagger}(x_1 ) {\phi}(x_4 ) \rangle,
\label{eq:doubdetdirect}
\end{eqnarray}
and the {\em exchange} term is given by
\begin{eqnarray}
E^{\rm ab}_{\pi\pi} ({\vec k}\, , {\vec k}'\, )
& = &
\int\! dx_1 dx_2 dx_3 dx_4 \;
{\cal S}^{\rm a}_{{\vec k}} (x_2 \, x_3 ) \,
{\cal S}^{\rm b}_{{\vec k}'}(x_1 \, x_4 ) \;
\langle \phi^{\dagger}(x_1 ) {\phi}(x_3 ) \rangle
\langle \phi^{\dagger}(x_2 ) {\phi}(x_4 ) \rangle \, .
\label{eq:doubdetexchange}
\end{eqnarray}
The direct term is the product of the single-pion detection
probabilities, Eq.~(\ref{eq:piexact0}),
\begin{equation}
D^{\rm ab}_{\pi\pi} ({\vec k}\, , {\vec k}'\, ) =
P^{\rm a}_{\pi} ({\vec k} \, )
P^{\rm b}_{\pi} ({\vec k}'\, ) \, ,
\label{eq:dppspd2}
\end{equation}
and is equivalent to the probability of observing two like-charge pions
from two different collision events, one in state ${\vec k}$ at ${\vec a}$ and
one in state ${\vec k}'$ at ${\vec b}\,$.  The exchange term is the HBT
effect.

    Knowledge of the mixture of pion states produced by the source is critical
to understanding HBT interferometry.  For pions produced in single particle
states $\varphi_{i}(x)$ the single-pion correlation function is
\begin{equation}
\langle \phi^{\dagger}(x_1 ) {\phi}(x_2 ) \rangle =
\sum_{i} F^{\phantom{*}}_{i} \;
\varphi^{*}_{i}(x_1 ) \,\varphi^{\phantom{*}}_{i}(x_2 ) \, ,
\label{eq:spdfwvfunc0}
\end{equation}
where $F_{i}$ specifies how the distribution of pion states
depends on the evolving geometry of the particle production region.
[Generally, the decomposition of the single pion correlation function as a sum
of single particle states defines the single particle states.]
The corresponding pion source current-current correlation function is
\begin{equation}
\langle J^{\dagger}(x_1) J(x_2) \rangle =
\sum_{i} F^{\phantom{*}}_{i} \;
{\cal J}^{*}_{i}(x_1 )\, {\cal J}^{\phantom{*}}_{i}(x_2 )\, ,
\label{eq:toy1wvfunc}
\end{equation}
where ${\cal J}_{i}(x )$ is the transition matrix element
of the pion source operator $J(x)\,$.  To illustrate the structure concretely,
we approximate the sources of the pions produced in a high-energy collision as
Gaussians in space and time
\begin{equation}
{\cal J}_{i}(x) =
\frac{N}{(2\pi)^2 \tau_{\rm c} R^3_{\rm c}}
e^{-ip\cdot (x - x_0 )} e^{-(t - t_{0})^{2}/2\tau^{2}_{\rm c}}
e^{-({\vec r} - {\vec r}_{0})^{2}/2 R_{\rm c}^2},
\label{eq:wvfunctransj10}
\end{equation}
where $R_{\rm c}$ and $\tau_{\rm c}$ are the characteristic length and
time scales for pion formation, and $N$ is a normalization constant;
here the subscript $i$ stands for the central momentum $\vec p$ and the
space-time origin of the wave $x_0$.
It has been shown elsewhere~\cite{PION:HBT4,BAYM:zak,J:Popp} that the
effective relative momentum scale of the pair correlation
function involves not only the dimensions and lifetime of the particle
production region and the mixture of states produced but also the individual
particle formation length and time scales.
These sources give rise to pions in
Gaussian wave packets, which in the far field have the form
\begin{equation}
\varphi^{\phantom{*}}_{i}(x ) \approx
-\frac{1}{8\pi^2 r} \; \int_{m_{\pi}}^{\infty} \! d\varepsilon_{q} \;
{\cal J}_{i}(\varepsilon_{q}\, , q{\hat r}\, ) \;
e^{-i\varepsilon_q t + iqr} \, ,
\label{eq:wvfuncsp10}
\end{equation}
where ${\cal J}_{i}(\varepsilon_{q}\, , q{\hat r}\, )$ is the
four-dimensional Fourier transform of ${\cal J}_{i}(x )\,$.
Equation~(\ref{eq:wvfuncsp10}) follows from Eq.~(\ref{eq:piplusff}) by
selecting the outgoing wave after integrating over momentum directions.

    For ${\vec p} = p {\hat z}\,$, the Gaussian source gives rise to a wave
packet which spreads out approximately as
\begin{eqnarray}
      \langle {\vec r}^{\, 2}_{\!\!\perp} \rangle
     & = &
       R_{\rm c}^2 + \frac{t^2}{(R_{\rm c}\varepsilon_{p} )^2}
\label{eq:flucperp0}
\\
\langle (z - v_{p} t)^2 \rangle
& = &
\frac{1}{2} \left( R_{\rm c}^2 + v_{p}^{2} \tau_{\!{\rm c}}^2 \right)
+ \frac{t^2}{2\varepsilon_{p}^2
\left( R_{\rm c}^2 + v_{p}^{2} \tau_{\!{\rm c}}^2 \right) \gamma_{p}^4} \, ,
\label{eq:flucpara0}
\end{eqnarray}
where $\gamma_{p} = {\varepsilon_{p}}/m_{\pi}$ and $v_p = |{\vec p}\, | /
\varepsilon_{p}\,$.  For $t \gg \tau_{\!{\rm c}}\,$, the ratio of the
longitudinal to the transverse spread of the wave function is $\sim
1/\gamma_{p}^2\,$.  Far from the source, relativistic pion momentum wave
packets are pancake-shaped and move in a direction perpendicular to the face
of the pancake.  Consider, e.g., a $1\,$GeV pion and assume that $\tau_{\rm
c}\sim R_{\rm c} \sim 1\,$fm.  For a source-detector distance $L\sim 1\,$m
the size of the wave packet envelope at the detectors in the transverse
direction is $\sim 20\,$cm and is $\sim 0.2\,{\rm cm}$ thick in the
longitudinal direction.

    Substituting Eq.~(\ref{eq:spdfwvfunc0}) into Eqs.~(\ref{eq:doubdetdirect})
and (\ref{eq:doubdetexchange}), we find the direct and exchange terms,
\begin{eqnarray}
D^{\rm ab}_{\pi\pi}({\vec k}\,{\vec k}'\,)
& = &
\sum_{ij} F_{i} F_{j} \;
{\cal W}^{\rm a}_{\vec k}(ii)^* \;
{\cal W}^{\rm b}_{{\vec k}'}(jj)
\label{eq:dirpidd1}
\\
E^{\rm ab}_{\pi\pi}({\vec k}\,{\vec k}'\,)
& = &
\sum_{ij} F_{i} F_{j} \;
{\cal W}^{\rm a}_{\vec k}(ij)^* \;
{\cal W}^{\rm b}_{{\vec k}'}(ij) \, ,
\label{eq:exchpidd1}
\end{eqnarray}
where we define the wave function-detector overlap function
\begin{equation}
{\cal W}^{\rm a}_{\vec k}(ij) \equiv
\int\! dx dx' \; \varphi^{*}_{i}(x) \,
{\cal S}^{\rm a}_{\vec k}(x \, x' ) \,
\varphi^{\phantom{*}}_{j}(x' ) \, .
\label{eq:ddint1}
\end{equation}
Each possible pair of pion waves $(ij)$ contributes
to the direct term, with one detected at ${\vec a}$ and the other at ${\vec
b}\,$.  However, only pairs of waves that arrive together at each detector
atom, within the time interval set by the detector functions ${\cal S}\,$
contribute to the exchange term.  Interferometry occurs when the wave
functions of each of the particles overlap in each of the detectors at the
same times.  Equation~(\ref{eq:exchpidd1}), with (\ref{eq:dirpidd1}), shows
that there is no restriction on the time interval between the arrivals at the
two atoms.  Correlations are measured over the time scale imposed by the
response of the atoms, Eq.~(\ref{eq:ddint1}), the time scale over which the
atoms are sensitive to the {\it phase} and {\it amplitude} of the incoming
particle waves.

    We illustrate the physics of detecting correlations with a simplified
model of the effective atomic current-current correlation function in
Eq.~(\ref{eq:specfunc}), which takes into account the main features of the
dynamic response of atoms to ionization by relativistic charged
particles~\cite{J:Popp}, namely, we assume that the space and time dependence
of the effective atomic current-current correlation function factorizes as
\begin{equation}
\langle {\stackrel{\,\sim}{j}}^{\dagger}_{\!\rm a}(x_1)
{\stackrel{\,\sim}{j}}^{\phantom{\dagger}}_{\!\rm a}(x_2) \rangle = f_{\rm
a}({\vec r}_{1}\, {\vec r}_2 \, )\, g( t_1\, t_2 ) \, .
\label{eq:jjcorrfact}
\end{equation}
Since the statistical distribution of atomic electron states in a gas
under normal conditions is essentially time-independent, then $g( t_1\, t_2 )
= g(t_{1} - t_{2})\,$.  The Fourier transform of $g(t)$ is the atomic
energy-absorption spectrum, and is determined from the distribution for
energy-loss per ionization for a relativistic charged particle passing through
a gas.  In a monatomic gas, this distribution is approximately Gaussian with a
long tail extending up to the kinematic limit for energy transfer,
$2m_{e}(\beta\gamma)^2$.  The majority of ionizing collisions occur within the
Gaussian part of the distribution and it is particularly the ionization events
in which the free electron carries away a minimal amount of kinetic energy
that are important for tracking.  Thus, we can neglect the high-energy tail
and select a (normalized) Gaussian energy spectrum for $g$:
\begin{equation}
g(t) =
\int\!\frac{d\omega}{2\pi} \; \tilde{g} (\omega ) \, e^{-i\omega t},
\label{eq:timeg1}
\end{equation}
where $ \tilde{g} (\omega ) = \sqrt{2\pi}\,
\zeta^{-1} \; e^{- (\omega - \omega_0)^2 /2\zeta^2 }\, $. The average
energy loss, $\omega_0\,$, for creating electron-ion pairs in, e.g.,
noble gases, is $10 - 50\,$eV~\cite{T:Ferbel}.  The energy bandwidth of the
Gaussian part of the distribution, $\zeta$, is $\sim 10\,$eV, an energy on the
scale of the average ionization potential per electron in a Thomas-Fermi model
of an atom.  This scale determines the characteristic time for ionization,
$\tau_{\rm atomic} = 1/\zeta \sim 10^{-16}\,$sec.  The spatial scale of the
atomic correlation function is determined by the sizes of the atomic electron
wave functions, characteristically $\sim 1\,$\AA; in terms of momentum, if the
kinetic energy picked up by an electron is a few tens of eV then the electron
momentum, $|{\vec q}\,|\,$, is on a ${\rm keV}/c$ scale, corresponding to
a distance $1/|{\vec q}\,| \sim 1\,$\AA, a result roughly consistent with the
size of impact parameters required in a classical Weizs\"acker-Williams
picture of a collision.
For convenience, we model $f_{\rm a}({\vec r}_{1}\, {\vec r}_2 \, )$ as
the product of two Gaussian functions centered on the atomic nucleus:
\begin{equation}
f_{\rm a}({\vec r}_{1}\, {\vec r}_2 \, ) \equiv f_{\rm a} \;
e^{-({\vec r}_1 - {\vec a}\,)^2 / 2R_{\rm a}^2 } \;
e^{-({\vec r}_2 - {\vec a}\,)^2 / 2R_{\rm a}^2 } \, ,
\label{eq:jjspacespdet}
\end{equation}
where $R_{\rm a}$ is of order angstroms and $f_{\rm a}$ is a constant.

    The delays between the emissions of pions directly produced in a
high-energy nucleus-nucleus collision are no larger than the lifetime of the
source.  However those produced in resonances can have a longer spread in
emission times.  As we now show, Hanbury Brown-Twiss correlations are
insensitive to the delays in arrival times at the detector atoms generated by
emission delays much smaller than the atomic response time, $\tau_{\rm
atomic}$.  With the use of the far-field form of the wave functions for
directly produced pions, Eq.~(\ref{eq:wvfuncsp10}), with
Eq.~(\ref{eq:wvfunctransj10}), the space and time integrations in
Eq.~(\ref{eq:ddint1}) are trivial.  Integrating over the energy components of
both pion wave packets we find the intermediate form,
\begin{eqnarray}
{\cal W}^{\rm a}_{\vec k}(ij)
& = &
c_{a} \int\! \frac{d\omega}{2\pi}\; \tilde{g}(\omega) \,
e^{-i(\varepsilon_{k} + \omega )
(t^{\phantom{\prime}}_0 - t^{\prime}_0)}\;
\nonumber \\
&   &
\phantom{\frac{e^4 f_{\rm a} R_{\rm a}^6} {4a^2 \varepsilon_{k} V}} \times
e^{-( {\vec k} - q{\hat a}\, )^2 R_{\rm a}^2 }\; e^{iq{\hat a}\cdot
({\vec r}^{\phantom{\prime}}_0 - {\vec r}^{\,\prime}_0 )}
{\cal J}^{*}_{{\vec p}}(\varepsilon_{k} +\omega\, , q{\hat a}) \;
{\cal J}^{\phantom{*}}_{{\vec p}\,'}(\varepsilon_{k} +\omega\, , q{\hat a})
\, ,
\label{eq:ddint1inter}
\end{eqnarray}
with $c_{a} = \pi e^4 f_{\rm a}^{\phantom{6}} R_{\rm a}^6 / 4a^2
\varepsilon_{k} V$; $|{\vec q}\,|$ is determined by the condition
$\varepsilon_{q} = \varepsilon_{k} + \omega$, and $t^{\phantom{\prime}}_0 -
t^{\prime}_0$ is the emission delay time between the pion waves.  Since as one
sees from the singles distribution~\cite{E802:Akiba}, the transition matrix
elements vary over an MeV scale or more, and $\tilde{g}(\omega )$ restricts
$\omega$ to a neighborhood of size $\zeta$ about $\omega_0\,$, both on an eV
scale, then for pions of energy at least one GeV, it is a very good
approximation to replace $q$ with $k$ and neglect $\omega$ everywhere in the
second line of Eq.~(\ref{eq:ddint1inter}).  Consequently, the first term on
the second line of Eq.~(\ref{eq:ddint1inter}) requires that ${\hat k} \cdot
{\hat a} \approx 1$ (to within one part in at least $10^{10}$), so that ${\vec
k} = k {\hat a}\,$.  The integral over $\omega$ is then simply the Fourier
transform of $\tilde{g}\,$:
\begin{equation}
g(t^{\phantom{\prime}}_0 - t^{\prime}_0 ) = e^{-i\omega_0
(t^{\phantom{\prime}}_0 - t^{\prime}_0 )}
e^{-(t^{\phantom{\prime}}_0 - t^{\prime}_0 )^2 \zeta^2/2 } \, ,
\label{eq:delayintdd}
\end{equation}
and the overlap function is
\begin{equation}
{\cal W}^{\rm a}_{\vec k} (ij) \approx
c_{a} g(t^{\phantom{\prime}}_0 - t^{\prime}_0 )\;
{\cal J}^{*}_{{\vec p}\, x^{\phantom{\prime}}_0}
(\varepsilon_{k}\, , {\vec k}\, ) \;
{\cal J}^{\phantom{*}}_{{\vec p}\,'_{\phantom{0}}\! x'_0}
(\varepsilon_{k}\, , {\vec k}\, ) \, .
\label{eq:result}
\end{equation}
The direct term is independent of $g$, since $t^{\phantom{\prime}}_0 =
t^{\prime}_0$ and ${\vec p} = {\vec p}\,'$ in Eq.~(\ref{eq:result}).  In fact,
Eqs.~(\ref{eq:delayintdd}) and (\ref{eq:result}) show that when the emission
delay time between pions is much less than $1/\zeta = \tau_{\rm atomic}$ we
may also neglect the time dependence of the detection process in the exchange
term; to a very good approximation we may then write $|g(t_{0} - t_{0'})| \sim
|g(0)| = 1\,$.  Substituting Eq.~(\ref{eq:result}) into
Eqs.~(\ref{eq:dirpidd1}) and (\ref{eq:exchpidd1}) and summing over the mixture
of wave packet states we find that the pair momentum detection probability is
proportional to the Fourier transform of Eq.~(\ref{eq:fact}),
\begin{equation}
P^{\rm ab}_{\pi\pi} ({\vec k}\, , {\vec k}'\, ) =
 c_{a} c_{b}
 \left[\langle J^{\dagger}(k)  J(k)  \rangle
 \langle J^{\dagger}(k') J(k') \rangle \; +
 |\langle J^{\dagger}(k) J(k') \rangle |^2 \right],
\label{eq:ddpipiprob}
\end{equation}
the naive result obtained by neglecting the time dependence of the
detection process.  However, in situations leading to much longer time delays,
in particular, pion emission from long-lived resonances compared with direct
pion production, one cannot necessarily neglect the time dependence of the
detection process.\footnote{Effects of time delays in propagation of wave
packets can also enter in the observation of neutrino oscillations, as noted
by Kim~\cite{CW:Kim}, and references therein.}

    Consider interferometry between a $\pi^{-}$ produced directly in a
heavy-ion collision and from the decay of, say a lambda, $\Lambda \rightarrow
\pi^{-} + p$, produced in the same reaction.  Because the $\Lambda$ moves more
slowly than a directly produced pion of the same rapidity as the one emitted
in the decay, the pion from decay will lag the directly produced one by a time
$\delta$ at the detector atoms.  To estimate this arrival time delay, we note
that a $\pi^{-}$ emitted in the forward direction has rapidity $y_{0} \approx
0.67$ in the $\Lambda$ rest frame, and that a $\Lambda$ of rapidity $y$
travels on average a distance $\tau_{\Lambda} \sinh y$ before decaying, where
$\tau_{\Lambda}$ is the $\Lambda$ lifetime.  Thus, $\delta = \tau_{\Lambda} /
(\cosh y + \sinh y / \tanh y_0 )$, which for a $\Lambda$ of typical rapidity
$3$ is $\sim 0.037\tau_{\Lambda} = 9.7\times 10^{-12}\,$sec, much longer than
the atomic time scale.  Pions emitted in other than the forward direction will
have an even greater time lag.  As we shall see, detector atoms are sensitive
to such delays when we measure pair correlations.

    The reduction in the probability for detecting pairs due to time delays
between direct and resonance decay pions can be readily estimated using a
simplified scalar-field model to compute the overlap, Eq.~(\ref{eq:ddint1}),
with a detector atom of the wave functions from a direct-production pion and a
pion from, say, lambda decay.  The wave mechanical features of the exact
problem do not depend on the details of the model.  Consider the interaction
${\cal H}_{I}(x) = \alpha\,\pi^{\dagger}(x) {\bar p}(x) \Lambda(x)$, where
$\alpha$ is the coupling constant.  The transition matrix element for the
decay-product pion is
\begin{equation}
{\cal J}^{\Lambda}_{\pi}(x) = \alpha\,\psi^{*}_{p}(x)
\varphi^{\phantom{*}}_{\Lambda}(x)\, ,
\label{eq:ressource1}
\end{equation}
where $\varphi^{\phantom{*}}_{\Lambda}$ is the wave function of the $\Lambda$
baryon of width $\Gamma \sim 10^{-6}\,{\rm eV}$, and $\psi^{\phantom{*}}_{p}$
is a plane wave state for the proton with momentum ${\vec p}$.  As before, we
write the wave functions of the lambda and both pions in the far-field form,
Eq.~(\ref{eq:wvfuncsp10}).  The Fourier transform of Eq.~(\ref{eq:ressource1})
is
\begin{equation}
{\cal J}^{\Lambda}_{\pi}(\varepsilon\, , {\vec q}\,) =
\frac{\alpha}{2\varepsilon^{\Lambda}_{{\vec q} + {\vec p}} \,
\sqrt{2\varepsilon^{p}_{\vec p} V}}\,
\frac{{\cal J}_{\Lambda}(\varepsilon + \varepsilon^{p}_{\vec p} ,
{\vec q} + {\vec p}\, )}{(\varepsilon + \varepsilon^{p}_{\vec p}
-\varepsilon^{\Lambda}_{{\vec q} + {\vec p}} ) + i\Gamma /2} \, ,
\label{eq:ressource2}
\end{equation}
where ${\cal J}_{\Lambda}$, the source function for lambda production, is
similar in structure to Eq.~(\ref{eq:wvfunctransj10}),
$\varepsilon^{\Lambda}_{{\vec q}} = ({\vec q}^{\, 2}_{\phantom{\Lambda}} +
m_{\Lambda}^2 )^{1/2}$, and $\varepsilon^{p}_{\vec p} = ({\vec p}^{\,
2}_{\phantom{p}} + m_{p}^2 )^{1/2}$.  We subsume the origin and momentum
labels of the wave packets into the subscript on the transition matrix
element, ${\cal J}_{\Lambda}$.  Thus, the wave function overlap at a detector
atom located at $\vec a$ with pion final state ${\vec k}$ is
\begin{equation}
{\cal W}^{\rm a}_{\vec k} (\pi\, \Lambda\, {\vec p}\, ) \approx
c_{a} \frac{\alpha\, {\cal J}^{*}_{\pi}
(\varepsilon^{\pi}_{\vec k}\, , {\vec k}\,) {\cal J}^{\phantom{*}}_{\Lambda}
(\varepsilon^{\pi}_{\vec k} +\varepsilon^{p}_{\vec p}\, ,
{\vec k} + {\vec p}\,)}{2\varepsilon^{\Lambda}_{{\vec k} +
{\vec p}}\, \sqrt{2\varepsilon^{p}_{\vec p} V}}\,
\int \frac{d\omega}{2\pi} \frac{\tilde{g} (\omega )}{(\omega +
\varepsilon^{\pi}_{\vec k}
+ \varepsilon^{p}_{\vec p} -\varepsilon^{\Lambda}_{{\vec k} + {\vec p}} )
+ i\Gamma /2} \, .
\label{eq:ressource3}
\end{equation}
The overlap function ${\cal W}^{\rm b}_{{\vec k}'} (\pi\, \Lambda\,
{\vec p}\, )$ for an atom at $\vec b$ is exactly the same as above,
but with ${\vec k}'$ in place of ${\vec k}$.

    The contribution to the exchange term, given by
\begin{equation}
P_{\pi\,\pi (\Lambda )}^{\rm exch}({\vec k}\, , {\vec k}'\, ) =
\sum_{\pi\, \Lambda\, {\vec p}} F_{\pi} F_{\Lambda}\;
{\cal W}^{\rm a}_{\vec k} (\pi\, \Lambda\, {\vec p}\, )^*\;
{\cal W}^{\rm b}_{{\vec k}'} (\pi\, \Lambda\, {\vec p}\, ),
\label{eq:pilambda}
\end{equation}
decreases as the width $\Gamma$ of the lambda decreases, i.e., as the
lifetime of the lambda increases.  We can see this explicitly by selecting a
convenient form for the energy absorption spectrum of the atom:
\begin{equation}
\tilde{g} (\omega ) = \frac{\zeta}{(\omega - \omega_0)^2 + (\zeta/2)^2} \, ,
\end{equation}
where $\zeta \sim 10\,$eV (cf.~Eq. (\ref{eq:timeg1})).
The effect is simplest to see for ${\vec k} = {\vec k}'$ (we report
the calculation for ${\vec k} \neq {\vec k}'$ in a later paper); then the sum
in Eq.~(\ref{eq:pilambda}) is
\begin{equation}
\sim \sum_{\pi\, \Lambda} F_{\pi} F_{\Lambda}\;
\int\! \frac{d^3 p}{(2\pi)^3 2\varepsilon^{p}_{\vec p}}
\frac{|{\cal J}^{*}_{\pi} (\varepsilon^{\pi}_{\vec k}\, , {\vec k}\,)
{\cal J}^{\phantom{*}}_{\Lambda} (\varepsilon^{\pi}_{\vec k} +
\varepsilon^{p}_{\vec p}\, , {\vec k} + {\vec p}\,)|^2}
{(2\varepsilon^{\Lambda}_{{\vec k} +
{\vec p}}\,)^2 } \, \frac{1}{( \omega_0 + \varepsilon^{\pi}_{\vec k}
+ \varepsilon^{p}_{\vec p} -\varepsilon^{\Lambda}_{{\vec k} + {\vec p}}\, )^2
+ (\zeta + \Gamma )^2 /4} \, .
\label{eq:wwlorentz}
\end{equation}
Since the factor containing $\Gamma$ varies over a much smaller energy
scale than that of the rest of the integrand we can replace the last term in
(\ref{eq:wwlorentz}) with $ 2\pi \delta (\varepsilon^{\pi}_{\vec k}
+\varepsilon^{p}_{\vec p} - \varepsilon^{\Lambda}_{{\vec k} + {\vec p}}\, )
/(\zeta + \Gamma )\,$, where we have also neglected $\omega_0$.
It is convenient to assume that the $\Lambda$ is produced in a spherically
symmetric state in the lab, so that the Fourier transform of
${\cal J}^{\phantom{*}}_{\Lambda}$ only depends on energy.
The width of the decaying
particle to lowest-order in our model interaction is $\Gamma = \alpha^2 p_{0}
/ 8\pi m^2_{\Lambda}\,$, where $p_0 = ((m^2_x / 2m_{\Lambda})^2 -
m^2_p )^{1/2}\,$, and $m^2_x = m^2_{\Lambda} + m^2_p - m^2_{\pi}\,$.  Thus,
integrating over proton momentum we find
\begin{equation}
P_{\pi\,\pi (\Lambda )}^{\rm exch}({\vec k}\, , {\vec k}\, ) =
\frac{\Gamma}{\Gamma + \zeta} {\cal P}_{\pi\,\pi (\Lambda )}^{\rm exch}({\vec
k}\, , {\vec k}\, )
\label{eq:result2a}
\end{equation}
where
\begin{equation}
{\cal P}_{\pi\,\pi (\Lambda )}^{\rm exch}({\vec k}\, , {\vec k}\, ) \equiv
c_{a} c_{b}
% \frac{\Gamma}{\Gamma + \zeta}
\frac{m^2_{\Lambda}}{2p_0k}
\sum_{\pi\, \Lambda} F_{\pi} F_{\Lambda}\;
|{\cal J}^{*}_{\pi} (\varepsilon^{\pi}_{\vec k}\, , {\vec k}\,)
{\cal J}^{\phantom{*}}_{\Lambda} (\varepsilon^{\Lambda}_{q_0}\, ,
q_0\,)|^2
\ln\big|\frac{\varepsilon^{\Lambda}_{q_0}+\beta_\pi q_0}
{\varepsilon^{\Lambda}_{q_0}-\beta_\pi q_0}\big|,
\label{eq:result2b}
\end{equation}
with $q_0 = p_0m_{\Lambda} / m_{\pi}$, $\beta_{\pi} \gamma_{\pi} = |{\vec
k}\,|/m_{\pi}$, and replacing ${\cal J}_{\Lambda}$ with an average matrix
element.  Equation~(\ref{eq:result2b}) is the probability obtained when the
atomic response time is neglected, that is, when one takes $\tilde{g} (\omega
) = 2\pi \delta(\omega - \omega_{\rm o})$.  Equation~(\ref{eq:result2a}) shows
that pions from the decay of long-lived resonances can lead to arrival time
delays at the detector atoms, relative to direct-production pions, that do
indeed give a reduced contribution to the HBT signal.  Thus, when $\Gamma \ll
\zeta$ the $\pi\Lambda$ exchange term is reduced by a factor $\Gamma /\zeta$
(for lambda-decay pions, $\sim 10^{-7}$) from the naive result and for
short-lived resonances with $\Gamma \gg \zeta$ we recover the case where we
neglect the detector response time.  Among the common weak decays that produce
pions the detection time effect is strongest for interferometry with charged
$\pi$'s from mesons with $\Gamma < 10^{-7}$eV; for $K^{\pm}$, $K^{0}_{L}$,
$\Gamma/\zeta < 10^{-8}$.  We expect a weaker suppression for pairs from the
shorter-lived particles where $\Gamma < 10^{-5}$eV; for $K^{0}_{S}$,
$\Lambda$, and $\Sigma^{\pm}$, $\Gamma/\zeta < 10^{-6}$.  For rarer decays
like $\Xi^{\pm}$, $\Omega^{-}$, and heavier flavors with
$\Gamma\,\raisebox{-0.1 cm}{$\stackrel{<}{\sim}$}\, 10^{-3}$eV the detection
time suppression factors are less than $10^{-4}$.

The full exchange term, of which Eq.~(\ref{eq:result2a}) is a part, includes
all possible pion pairs which involve contributions of the type:
$({\cal J}^{*}_{\pi},
{\cal J}^{\phantom{*}}_{{\pi}'})$, $({\cal J}^{*}_{\pi}, {\cal
J}^{\phantom{*}}_{{\Lambda}'})$, $({\cal J}^{*}_{\Lambda}, {\cal
J}^{\phantom{*}}_{{\pi}'})$, and $({\cal J}^{*}_{\Lambda}, {\cal
J}^{\phantom{*}}_{{\Lambda}'})$.  The $({\cal J}^{*}_{\pi}, {\cal
J}^{\phantom{*}}_{{\pi}'})$ term is computed in Eq.(\ref{eq:result}).  The
terms $({\cal J}^{*}_{\pi}, {\cal J}^{\phantom{*}}_{{\Lambda}'})$ and $({\cal
J}^{*}_{\Lambda}, {\cal J}^{\phantom{*}}_{{\pi}'})$ are identical and
contribute a factor 2 times Eq.~(\ref{eq:result2a}) to the full exchange term.
The $({\cal J}^{*}_{\Lambda}, {\cal J}^{\phantom{*}}_{{\Lambda}'})$ term may
be computed in a way similar to that of Eq.~(\ref{eq:result2a}), giving an even
smaller contribution to the HBT effect.
The essential result is that the response time of the detector atoms is
sensitive to phase differences between the waves due to arrival time delays at
the detectors; a result independent of the origin of the waves.  One can see
this, quite simply, by introducing an explicit time delay in the wave funtions
at the detectors in Eq.(\ref{eq:ddint1}).

\section{SUMMARY AND CONCLUSIONS}

    We have developed, within an elementary quantum mechanical framework for
computing correlation measurements in high-energy experiments, a general
description of how detectors probe many-particle wave functions.  The HBT
effect is the consequence of wave mechanics performed by particle detectors
and depends only on the wave functions of the particles at the sensitive
elements of the detectors; it is not caused by stimulated emission or any
other mechanism at the source and it does not depend on the history of the
particles, e.g., the particles do not have to have a common origin.

    We have studied how momentum correlations between pairs of particles are
detected via the HBT effect.  We have shown that:  1) The like-pair
correlation function is able to reveal momentum correlations because the
single-particle wave functions of the detected particles overlap at at least
two sensitive elements of a detector, within the characteristic atomic
transition times of those elements. 2) There is no restriction on the time
interval between the transitions of the two detector atoms. 3) The size of the
measured pair correlation function is reduced when the delay in arrival times
between pairs at the detectors is of order of or larger than the transition
time; e.g., delays from particles produced in very long-lived resonance
decays.

\section*{Acknowledgements}
    We are grateful to Benoit Vanderheyden and Barbara Jacak for discussions.
This work was supported in part by National Science Foundation Grants No.
PHY94-21309 and PHY98-00978.

\section*{Appendix A: Momentum detection probability
for single charged pions}
\hyphenation{cor-relations}
\hyphenation{detec-tor}
\hyphenation{func-tion}
\hyphenation{Feyn-man}

    We compute here the probability for an energetic charged pion to ionize a
gas atom and emerge in a final plane-wave state with momentum ${\vec k}\,$.
We denote the initial state by $|{\cal I}\rangle = |I,i_{a},0\rangle$, where
$I$ is the inital pion state, and $|i_{a}\rangle$ is the initial state of the
atomic electrons; and the final state by $|{\cal F}\rangle = |{\vec
k},\,f_{a}\rangle$ where $| f_{a} \,\rangle$ contains the same number of
electrons as $|i_{a}\rangle$, but with one electron in a continuum state.  For
simplicity, we work with single-pion states; the generalization to many-pion
states is straightforward.  The interaction of a charged pion with a
detector-gas atom located at a space point ${\vec a}$ is
\begin{equation}
{\cal H}^{\phantom{a}}_{I} = {\cal H}^{\pi}_{I} + {\cal H}^{\rm a}_{I} \, ,
\label{eq:piatomaqed}
\end{equation}
where $ {\cal H}^{\pi}_{I} = e j^{\mu}_{\pi} A_{\mu}$, $A_{\mu}$ is the
electromagnetic field, $j^{\mu}_{\pi}$ is the charged pion electromagnetic
current, ${\cal H}^{\rm a}_{I} = e j^{\mu}_{\rm a} A_{\mu}$, with $j_{\rm
a}^{\mu}$ the electron current of the detector atom and $-e$ the electron
charge; we work in the interaction picture.  For $\pi^{\pm}$, the
electromagnetic current is $ j^{\mu}_{\pi}(x) = \pm\phi^{\dagger}(x)
i{\stackrel{\!\!\leftrightarrow}{\partial^{\mu}}} \phi(x)$.

    The lowest-order contribution to the amplitude for ionization in a
collision between an energetic charged pion and an atom comes from the second
order terms in the matrix element of the time evolution operator, $U(t,t')$,
\begin{equation}
 {\cal{A}}^{\rm a}_\pi({\cal I}\to{\cal F})
 = - \int dx_1 dx_2\langle {\cal F}| T \left[ {\cal H}^{\rm a}_{I}(x_1)
    {\cal H}^{\pi}_{I}(x_2)\right]|{\cal I}\rangle
 = -ie^2 \int dx_1 dx_2 \langle f_{\rm a}|
  j^{\mu}_{\rm a}(x_1)| i_{\rm a} \rangle D_{\mu\nu}(x_1 \, x_2 )
  \langle{\vec k}\,|j^{\nu}_{\pi}(x_2)|I\rangle.
\label{eq:onepionop2}
\end{equation}
Here $T$ denotes time-ordering; the time integrations are from $t$ to
$t'$, where the exposure time $\tau_{\rm exp}$ is $t' - t$, and the space
integrals are over all space.  In the latter expression we use the free-field
forms for the operators, and introduce the photon propagator,
$D_{\mu\nu}(x_1\, x_2 ) = -i \langle 0 |T\left[
A_\mu(x_1)A_\nu(x_2)\right]|0\rangle$, where $|0\rangle$ is the vacuum.

    The transition matrix element of the pion current becomes $\langle {\vec
k}\,|\phi^{\dagger}(x_2)i {\stackrel{\!\!\leftrightarrow}{\partial^{\nu}_{2}}}
\phi(x_2)|I\rangle = \psi^{*}_{\vec k} (x_2 ) i
{\stackrel{\!\!\leftrightarrow}{\partial^{\nu}_{2}}} \langle 0| {\phi}(x_2 )
|I\rangle$, where $\psi^{\phantom{*}}_{\vec k}(x) = (1/\sqrt{2\varepsilon_{k}
V}) e^{-ik\cdot{x}}$.  Since the energy spectrum of the atomic states
important for tracking via ionization only has Fourier components a few tens
of eV or less, the spectrum of the atomic function $\int dx_1 \langle f_{\rm
a}| j^{\mu}_{\rm a}(x_1)| i_{\rm a} \rangle D_{\mu\nu}(x_1,x_2 )$ is similarly
constrained.  Integrating in $x_2$ by parts in Eq.~(\ref{eq:onepionop2}) and
neglecting the derivatives of this function, we can replace $i
{\stackrel{\!\!\leftrightarrow}{\partial^{\nu}_{2}}}$ by $2k^{\nu}$.

    Squaring the transition amplitude, summing over all final electron states,
averaging over initial pion and electron states, we find the ionization
probability
\begin{equation}
P^{\rm a}_{\pi}({\vec k}\,) = 4e^4 \int dx_1 dx_2 dx_3 dx_4
   D^*_{\alpha\sigma}(x_3\,x_4 )
  \langle j_{\rm a}^{\alpha}(x_3 )j_{\rm a}^{\mu}(x_1)\rangle
    D_{\mu\nu}(x_1\,x_2 )
      \psi^{*}_{\vec k} (x_2 ) \psi_{\vec k} (x_4)
 k^{\sigma} k^{\nu}  \langle \phi^{\dagger}(x_4){\phi}(x_2)\rangle,
\label{eq:probsingpi2}
\end{equation}
where the single-pion density matrix is
\begin{equation}
\langle \phi^{\dagger}(x_4 ) {\phi}(x_2 ) \rangle = \sum_{I} \rho_{I}
\langle\, I | \phi^{\dagger}(x_4 ) {\phi}(x_2 ) | I\,\rangle,
\label{eq:sipidema}
\end{equation}
with $\rho_{I}$ the probability that the state $|I\rangle$ is produced by
the source; the electric current-electric current correlation function for the
atomic electrons is
\begin{equation}
\langle j_{\rm a}^{\alpha}(x_3) j_{\rm a}^{\mu}(x_1)\rangle
\equiv \sum_{i} \rho^{\rm a}_{i} \sum_{f}
\langle i_{\rm a}|
j_{\rm a}^{\alpha}(x_3)|f_{\rm a}\rangle
\langle f_{\rm a}| j_{\rm a}^{\mu}(x_1)|i_{\rm a}\rangle,
\label{eq:jjatom1}
\end{equation}
where $\rho^{\rm a}_{i}$ is the probability of finding the atom at ${\vec a}$
in the state $|i_{\rm a}\rangle$.  Defining an effective current operator,
\begin{equation}
{\stackrel{\,\sim}{j}}_{\!\!{\rm a}}(x_2 ) \equiv
  2k^{\nu}\int dx_1  j^{\mu}_{\rm a}(x_1) D_{\mu\nu}(x_1 \, x_2 ),
\label{eq:pseudxo}
\end{equation}
we see that the ionization probability reduces to
\begin{equation}
P^{\rm a}_{\pi}({\vec k}\,) = \int dx_1 dx_2
{\cal S}^{\rm a}_{\vec k}(x_1 \, x_2 )
\langle \phi^{\dagger}(x_1 ) {\phi}(x_2 ) \rangle,
\label{eq:piexact}
\end{equation}
where the spectrometer function ${\cal S}^{\rm a}_{\vec k}(x_1 \, x_2 )$
is defined in Eq.~(\ref{eq:specfunc}).

\section*{Appendix B: Momentum detection probability for charged pion pairs}
\hyphenation{cor-relations}
\hyphenation{detec-tor}
\hyphenation{Feyn-man}
\hyphenation{func-tion}

    In this Appendix we compute the probability for detecting a pair of
$\pi^+$: one with momentum ${\vec k}$ at ${\vec a}$ and one with
momentum ${\vec k}'$ at ${\vec b}$.  We denote the initial and final
states by $|{\cal I}\,\rangle = |I, i_{\rm a}, i_{\rm b}, 0\rangle$ and
$|{\cal F}\,\rangle = |{\vec k}, {\vec k}', f_{\rm a}, f_{\rm b}, 0\rangle$,
where the number of pions in the initial and final states are the same, and
${\vec k}$ and ${\vec k}'$ are the measured states.  The interaction
Hamiltonian for the system of detector atom-a and detector atom-b plus two
pions is ${\cal H}^{\phantom{a}}_{I} = {\cal H}^{\pi}_{I} +
{\cal H}^{\rm a}_{I} + {\cal H}^{\rm b}_{I}$, where the interactions are
given in Appendix~A.  Two-pion correlation measurements are given by the
term fourth-order in ${\cal H}_{I}$ in the time evolution operator, $U(t,t')$,
which with free fields becomes
\begin{equation}
\frac{e^4}{2} \int dx_1 dx_2 dx_3 dx_4 \,
j^{\mu}_{\rm a}(x_1 )
T\left[ j^{\nu}_{\pi}(x_2 ) j^{\sigma}_{\pi}(x_3 )\right]
j^{\kappa}_{\rm b}(x_4 )
T\left[ A_{\mu}(x_1 ) A_{\nu}(x_2 )
A_{\sigma}(x_3) A_{\kappa}(x_4)\right].
\label{eq:tranoppi23}
\end{equation}

    The matrix elements of the pion and photon operators are reduced as
follows: We write the pion current-current matrix element as
\begin{equation}
\langle {\vec k} , {\vec k}' |\,
j^{\nu}_{\pi}(x_2 ) j^{\sigma}_{\pi}(x_3 ) | I\rangle =
i {\stackrel{\!\!\leftrightarrow}{\partial^{\nu}_{2}}}
i {\stackrel{\!\!\leftrightarrow}{\partial^{\sigma}_{3}}}
\langle {\vec k} , {\vec k}' |
\phi^{\dagger}(x_2 ) {\phi}(x_2 )
\phi^{\dagger}(x_3 ) {\phi}(x_3 ) | I\rangle ,
\label{eq:jjpipi1}
\end{equation}
where
${\stackrel{\!\!\leftrightarrow}{\partial^{\nu}_{2}}}$ and
${\stackrel{\!\!\leftrightarrow}{\partial^{\sigma}_{3}}}$
act between the operators carrying the same variables.  Using the free-field
commutation relation $ \left[ \phi(x_2 ) , \phi^{\dagger}(x_3 )
\right] =  {\cal D}(x_2 \, x_3 )$, a c-number, we express the pion-current
operators in terms of the two-pion correlation operator plus a term
involving ${\cal D}(x_2 \, x_3 )$.  The matrix element
\begin{equation}
\langle {\vec k} , {\vec k}' |
\phi^{\dagger}(x_2 ) \phi^{\dagger}(x_3 ) {\phi}(x_3 ) {\phi}(x_2 ) |I\rangle
\label{eq:4phi}
\end{equation}
is exactly the interaction of two pions ionizing two atoms; we remove
two pions at two separate space-time points and replace them.
The commutator term ${\cal D}(x_2 \, x_3 )$ does not contribute to
lowest-order scattering, hence we replace the matrix element on the
right-hand-side of Eq.~(\ref{eq:jjpipi1}) with (\ref{eq:4phi});
whereupon we eliminate the time-ordering of the pion fields
in Eq.~(\ref{eq:tranoppi23}) using the symmetry of these expressions under the
interchange $x_2 \leftrightarrow x_3$ and $\nu \leftrightarrow \sigma$.
Since we require $\psi^{\phantom{*}}_{{\vec k}}$ at ${\vec a}$
and $\psi^{\phantom{*}}_{{\vec k}'}$ at ${\vec b}$, (\ref{eq:4phi})
finally becomes $\psi^{*}_{\vec k}(x_2 ) \psi^{*}_{{\vec k}'}(x_3 )
\langle 0|\phi(x_3 )\phi(x_2 ) |I\rangle$.  The vacuum expectation value of
the time-ordered photon operators in Eq.~(\ref{eq:tranoppi23}) factorizes as
\begin{eqnarray}
-\langle 0| T\left[ A_{\mu}(x_1 ) A_{\nu}(x_2) A_{\sigma}(x_3 )
A_{\kappa}(x_4 ) \right] |0\rangle  \hspace{215pt}
\nonumber \\
= D_{\mu\nu}  (x_1, x_2 ) D_{\kappa\sigma}(x_4,x_3 )
+ D_{\mu\sigma} (x_1, x_3 ) D_{\kappa\nu} (x_4,x_2 )
+ D_{\mu\kappa} (x_1, x_4 ) D_{\nu\sigma} (x_2,x_3 ).
\label{eq:wickpp}
\end{eqnarray}
The last term in Eq.~(\ref{eq:wickpp}) represents photon exchange between
detector atoms and is not important for ionization.

    After eliminating the crossed photon lines in the amplitude, and the
derivatives of the pion fields as in Appendix~A, we write
\begin{eqnarray}
{\cal {A}}^{\rm ab}_{\pi\pi}({\cal I}\rightarrow {\cal F} ) & = &
- e^4 \int\! dx_1 dx_2 dx_3 dx_4
\langle f_{\rm a} |\, j^{\mu}_{\rm a}(x_1 )    | i_{\rm a} \rangle
D_{\mu\nu}(x_1 \, x_2 )
\langle f_{\rm b} |\, j^{\kappa}_{\rm b}(x_4 ) | i_{\rm b} \rangle
D_{\kappa\sigma}(x_4 \, x_3 )
\nonumber \\
&   & \qquad \times
\psi^{*}_{\vec k}(x_2 ) \psi^{*}_{{\vec k}'}(x_3 )
4 k^{\nu} k'^{\sigma}
\langle 0| {\phi}(x_3 ) {\phi}(x_2 ) | I\rangle .
\label{eq:pipairamp2}
\end{eqnarray}
The transition probability is computed in exactly the same way as for single
pion detection; redefining
the electron currents, Eq.~(\ref{eq:pseudxo}), and expressing the momentum
measurements in terms of spectrometer functions, Eq.~(\ref{eq:specfunc}), we
derive:
\begin{eqnarray}
P^{\rm ab}_{\pi\pi} ({\vec k} , {\vec k}' ) & = &
\int\! dx_1 dx_2 dx_3 dx_4
{\cal S}^{\rm a}_{\vec k}(x_2 \, x_3 )
\langle \phi^{\dagger}(x_1 ) \phi^{\dagger}(x_2 )
{\phi}(x_3 ) {\phi}(x_4 ) \rangle
{\cal S}^{\rm b}_{{\vec k}'}(x_1 \, x_4 ).
\label{eq:doubdet2piprob}
\end{eqnarray}

\end{document}